\documentclass[showpacs,preprintnumbers,nofootinbib,
superscriptaddress,amsmath,floatfix,prd,12pt,tightenlines]{revtex4}

\usepackage{graphicx}
\usepackage{amsmath, amsfonts}

\usepackage[dvips]{color}

\newcommand{\be}{\begin{equation}}
\newcommand{\ee}{\end{equation}}
\newcommand{\bea}{\begin{eqnarray}}
\newcommand{\eea}{\end{eqnarray}}
\newcommand{\ba}[1]{\begin{array}{#1}}
\newcommand{\ea}{\end{array}}

\newcommand{\bel}[1]{\begin{equation}\label{#1}}
\newcommand{\beal}[1]{\begin{eqnarray}\label{#1}}

\def\nn{\nonumber\\ }
\def\0{\over } \def\1{\vec } \def\2{{1\over2}} \def\4{{1\over4}}
\def\5{\bar } 
\def\6{\partial }
\def\7#1{{#1}\llap{/}}
\def\8#1{{\textstyle{#1}}} \def\9#1{{\bf{#1}}}
\def\.{\cdot }
\def\^#1{\widehat{#1}}
\let\ph=\varphi

\def\CL{{\cal L}}

\def\({\left(} \def\){\right)} \def\<{\langle } \def\>{\rangle }
\def\[{\left[} \def\]{\right]}  

\newcommand{\wkl}{\omega_{k\ell}}
\newcommand{\wol}{\omega_{0\ell}}
\newcommand{\wbl}{\omega_{B\ell}}

\newcommand{\lambdad}{\lambda}

\begin{document}
\title{
One-loop results for kink and domain wall
profiles\\ at zero and finite temperature}

\preprint{TUW-09-02}
\preprint{YITP-SB-09-03}

\author{Anton Rebhan}\email{rebhana@hep.itp.tuwien.ac.at}
\affiliation{Institut f\"ur Theoretische Physik, Technische Universit\"at Wien,
        Wiedner Hauptstrasse 8-10, A-1040 Vienna, Austria}
\author{Andreas Schmitt}\email{aschmitt@hep.itp.tuwien.ac.at}
\affiliation{Institut f\"ur Theoretische Physik, Technische Universit\"at Wien,
        Wiedner Hauptstrasse 8-10, A-1040 Vienna, Austria}
\author{Peter van
Nieuwenhuizen}\email{vannieu@insti.physics.sunysb.edu}
\affiliation{C.N.Yang Institute for Theoretical
Physics, State University of New York, Stony Brook, NY 11794-3840 }

\date{July 3, 2009}


\begin{abstract}

Using dimensional regularization,
we compute the one-loop quantum and thermal corrections to the
profile of the bosonic 1+1-dimensional $\ph^4$ kink,
the sine-Gordon kink and the CP$^1$ kink, and higher-dimensional
$\ph^4$ kink domain walls. 
Starting from the
Heisenberg field equation in the presence of the 
nontrivial kink background 
we derive analytically results for the 
temperature-dependent mean field which display
the onset of the melting of kinks as the system is heated
towards a symmetry restoring phase transition.
{The result is shown to simplify significantly
when expressed in terms of a self-consistently defined
thermal screening mass.}
In the case of domain walls, we find infrared singularities
in the kink profile, which corresponds to interface roughening
depending on the system size.
Finally we calculate the energy density profile of $\ph^4$ kink domain walls
and find that in contrast to the total surface tension
the local distribution requires composite operator renormalization in 3+1 dimensions.
\end{abstract}

\maketitle

\section{Introduction}

Solitons \cite{Raj:Sol}
in scalar field theories in 1+1 dimensions, which we call generically kinks, have
played an important role in increasing our understanding of various nontrivial
aspects of quantum field theories, ranging from exactly solvable
examples of strong/weak-coupling dualities 
\cite{Dashen:1974cj,Dashen:1975hd}
to the theory of topological defects generated at phase transitions with
applications in condensed matter physics 
\cite{ZinnJustin:2002ru} as well as cosmology 
\cite{Kibble:1976sj}.

In a number of models one knows in closed form the spectrum of fluctuations
about the kink background which allows one to perform
complete calculations of one-loop corrections to the mass of the kinks.
However, these calculations turn out to be full of subtleties, in particular
(but not only)
in the presence of fermions. For example, for the minimally supersymmetric kink
a number of authors have concluded from explicit calculations that
there was a cancellation of the one-loop effects on mass and central
charge in a certain minimal renormalization scheme
\cite{Kaul:1983yt,Imbimbo:1984nq
,Yamagishi:1984zv}, a result widely accepted
since the mid 1980's. Only in
1997 two of the present authors have reopened this issue by demonstrating
an incompatibility of the methods employed for the supersymmetric kinks
with known exact results for the nonsupersymmetric
sine-Gordon model \cite{Rebhan:1997iv}, with correct results
for the mass eventually being established in 
Refs.~\cite{Nastase:1998sy,Graham:1998qq,Shifman:1998zy} and a resulting puzzle
concerning BPS saturation solved in Ref.~\cite{Shifman:1998zy} 
by the discovery of a new anomaly in the central charge\footnote{Actually the
conformal central charge \cite{Rebhan:2002yw}.}.
More recently this has led to a similar revision also in the case
of 4-dimensional supersymmetric monopoles \cite{Rebhan:2004vn,Rebhan:2006fg},
where a long-standing (20 years) but unnoticed discrepancy 
of direct calculations \cite{Kaul:1984bp,Imbimbo:1985mt}
with newer developments (notably Seiberg-Witten theory \cite{Seiberg:1994rs})
was eventually cleared up.\footnote{For more extensive reviews see
e.g.~\cite{Goldhaber:2004kn,Shifman:2007ce}.}

In the present paper we shall consider bosonic kinks at finite temperature%
\footnote{Finite temperature breaks supersymmetry so that supersymmetric
kinks do no longer display features that are not also found in the bosonic
case. Moreover, thermal contributions from fermions provide no special
difficulty and are easily added on to what follows.}, applying and
slightly generalizing
the method used in Ref.~\cite{Shifman:1998zy,Goldhaber:2001rp} 
to calculate the profile of kink to one-loop order.
Following Ref.~\cite{Goldhaber:2001rp} we start from the Heisenberg field
equation. Other authors have used the method of the effective action with
$x$-dependent background fields (see \cite{ZinnJustin:2002ru}
and references therein). 
Both methods are of course completely
equivalent but we found the former to be simpler to implement.

Quantum and thermal corrections to kinks have been considered
in various approximations of self-consistent, nonperturbative frameworks
e.g.~in Refs.~\cite{Boyanovsky:1998ka,Salle:2003ju,Bergner:2003au}. 
Here we shall restrict ourselves to
one-loop effects in a regime where perturbation theory is
still reliable, refraining therefore 
from attempts to cover the 
physics of the symmetry restoring phase transition itself and the
corresponding actual melting of kinks. However, we can reliably cover
the onset of the melting of kinks and in this way provide benchmarks
for more daring approximations and approaches. 
Systematic calculations of one-loop corrections to kink profiles and
domain walls at zero and finite temperature have been carried out
before by several authors \cite{Rudnick:1978zz,AragaodeCarvalho:1983sc,deCarvalho:2001da,Bessa:2004pu,Kopf:2008hr}. 
Our results are in agreement with those
once the differences in the renormalization schemes are taken
into account.\footnote{%
In Ref.~\cite{Rebhan:2002uk} two of us have previously stated that
we disagreed with Ref.~\cite{deCarvalho:2001da}
regarding results on the surface tension at zero temperature
in a zero-momentum renormalization scheme, but agreed
with Ref.~\cite{Munster:1989we}. As has been
shown in Ref.\ \cite{Bessa:2004pu}, this apparent discrepancy was due
to having compared two versions of a zero-momentum renormalization scheme.
The zero-momentum scheme considered in
Refs.~\cite{Munster:1989we,Rebhan:2002uk} renormalizes the
derivative of the two-point function with respect to momentum to one,
while the scheme in
Refs.~\cite{deCarvalho:2001da,Bessa:2004pu} does not introduce
nontrivial wave-function renormalization, which is possible at one-loop order.
Taking this difference into account resolves this issue.}
However, we find
significant simplifications in the corrections to the kink profile
provided renormalization conditions are formulated in terms of
a self-consistent thermal screening mass, which may be of practical 
importance in applications at non-infinitesimal coupling.


We shall in turn cover three different models in 1+1 dimensions: the well-known
exactly solvable sine-Gordon model with $\mathbb{Z}\times \mathbb{Z}_2$ symmetry, a closely related massive CP$^1$ model
with U(1)$\times \mathbb{Z}_2$ symmetry, and the familiar $\ph^4$ model 
with $\mathbb{Z}_2$ symmetry. 
All the above models have a discrete symmetry which is spontaneously broken
at zero temperature and which leads to topological solitons (kinks). 
Since the $\ph^4$ model is renormalizable
in higher dimensions as well, we shall use it also to discuss domain
walls in 2+1 and 3+1 dimensions.

At sufficiently high temperature $T$ one expects symmetry restoration
and the disappearance (``melting'') of the kinks. In 3+1 dimensions it is
well known that perturbation theory allows one to derive the leading order
result for the phase transition temperature $T_c$, but in simple 
scalar field theories
next-to-leading order results for the transition temperature as well as the order of the phase transition
are beyond perturbation theory \cite{Arnold:1992fb,Arnold:1993rz}.

Therefore, let us explain the validity and limitations of perturbative
methods in the various dimensions.
In 3+1 dimensions, symmetry restoration in $\ph^4$ theory
with coupling constant $\lambda$
is brought about by a thermal
mass term $\Delta m^2 \ph^2 \sim \lambda T^2 \ph^2$ which outweighs the
wrong-sign mass term $\propto -\mu^2 \ph^2$ in the classical potential
for sufficiently high temperature $T>T_c\sim\mu/\sqrt\lambda$.
After a resummation of the leading-order thermal mass, perturbation theory
around the minimum of the effective potential has a loop expansion parameter
$\sim \lambda T/m$, where $m$ is the (thermally corrected) mass 
of the fluctuations around the minimum, with
$m\sim \sqrt{2}\mu$ at low $T$ in the broken phase, and $m\sim \Delta m$ for large $T$ in the restored phase. 
With $\lambda\ll 1$, perturbation theory works for all temperatures except
very close to $T_c$ where $m$ gets parametrically small. (As long as
$m\sim\mu$, the expansion parameter $\lambda T/m \lesssim \sqrt\lambda$
up to temperatures of the order of $T_c$; for $T\gg T_c$, one has
$m\sim \sqrt\lambda\, T$ and the expansion parameter is of order $\sqrt\lambda$
throughout the symmetry-restored phase.)

In lower dimensions, the situation is much more dire.
In 1+1 dimensions, the coupling constant of $\ph^4$ theory as well
as of the sine-Gordon model has scaling dimension
mass squared, and we have to assume that
the loop expansion parameter at zero temperature
$\lambda/m^2\ll 1$. 
(For the CP$^1$ kink
we have a dimensionless coupling constant replacing $\lambda/m^2$.)
At finite temperature the expansion parameter
is $(\lambda/m^2)\times(T/m)$. High-temperature thermal mass terms are however
only linear in $T$, $\Delta m^2\sim\lambda T/m$. For symmetry restoration
we would need $|\Delta m^2| \gtrsim m^2$, but this contradicts
the requirement $(\lambda/m^2)\times(T/m)\ll 1$. 
Hence, in perturbation theory we can reliably study
the high-temperature limit $T/m \gg 1$ only as long as $T/m \ll m^2/\lambda$,
i.e.\ 
in the broken phase where $|\Delta m^2| \ll m^2$. It is therefore
not mandatory to resum the thermal mass $\Delta m$, although we shall
find that it will be natural to do so. 

An important difference to the
3+1-dimensional case is, however, that
$\Delta m^2 \sim \lambda T/m$ in the 1+1-dimensional theory is generated only
by Matsubara zero modes, whereas in the 3+1-dimensional case the
leading terms in $\Delta m^2$
are generated by ``hard'', short-distance modes with wavelength $\sim T^{-1}$, which for $T/m\gg1$ are insensitive
to the presence of a nontrivial kink background
with inherent length scale $\sim m^{-1}$.

In 2+1-dimensional theory, the situation is not better than in 1+1 dimensions.
The loop expansion parameter at zero temperature is $\lambda/m \ll 1$.
The thermal expansion parameter $\lambda T/m^2\ll 1$ equally implies
that the thermal mass squared $\Delta m^2 \sim \lambda T \ll m^2$,
precluding a perturbative analysis of a symmetry-restoring phase transition.
When going to high temperatures $T/m \gg 1$, perturbation theory is
reliable only as long as $T/m \ll m/\lambda$, so again only the
broken phase is accessible.\footnote{The critical temperature of the
2+1-dimensional $\varphi^4$ model has been estimated by renormalization-group
methods in Ref.~\cite{Einhorn:1992cb} 
to be given by a relation of the form 
$T_c/\mu \propto \mu/\lambda \log(c\lambda/T_c)$
and the constant $c$ subsequently measured
 on the lattice in Ref.~\cite{Bimonte:1996fy}.}

In order to study the actual melting of kinks, nonperturbative methods are
needed. (In the lower dimensional cases, also the symmetry-restored phase
with $T\gg T_c$ is nonperturbative.) In some cases, other systematic expansions like large-$N$
expansions \cite{Dine:1980jm,Andersen:2003va}
can be put to work, but often 
(self-consistent) approximation schemes are employed
which lack an expansion parameter controlling the approximations. 
In order to have credibility, such approximations should be able
to reproduce our perturbative results as limiting case.

In Sect.~\ref{sectSG} we consider the sine-Gordon
model and the massive CP$^1$ model, which have closely
related fluctuation spectra, and we calculate the
one-loop corrections to the field profile of the kinks
in these models. In Sect.~\ref{sectph} we turn to
the familiar $\ph^4$ model, which has a more
complicated fluctuation spectrum and correspondingly more
complicated one-loop corrections. Since the latter model is
renormalizable in higher dimensions, we shall use it to discuss
one-loop corrections to domain walls in 2+1 and 3+1 dimensions,
with a detailed discussion of 
our dimensional regularization and renormalization scheme, and
its appropriate generalization at finite temperature. In particular,
{with the help of a remarkable identity for integrals
involving the Bose-Einstein distribution function},
we find that a self-consistent definition of the thermal screening mass
removes certain artefacts in the one-loop kink profile.

In the calculation of the profile of the higher-dimensional
domain walls, we encounter
infrared singularities associated with the massless modes that correspond
to the translational zero mode of the 1+1-dimensional kink.
In accordance with Refs.~\cite{Rudnick:1978zz,Bessa:2004pu,Kopf:2008hr}, these
singularities are interpreted as the field theoretic equivalent
of system-size dependent interfacial roughening \cite{Buff:1965zz,Gelfand:1990,Muller:2004vv}. Finally we calculate the energy profile of the $\varphi^4$ kink
and the corresponding domain walls and show that in contrast to the total
mass (surface tension) the local energy density profile is ambiguous,
depending on improvement terms to the stress tensor, and that in
3+1 dimensions composite operator renormalization through improvement
terms is required.

\section{Sine-Gordon and CP$^1$ kinks}\label{sectSG}

We begin by calculating the one-loop quantum and thermal corrections
to the field profile of the 1+1-dimensional sine-Gordon model and
a massive version of the CP$^1$ model, which both happen to have
a simpler fluctuation spectrum than the $\varphi^4$ model (in particular,
no bound states).
A full-fledged discussion of our method of dimensional regularization will 
be introduced in Sect.~\ref{sectph}, where we turn to the $\varphi^4$ model
in 1+1 and higher dimensions, which also turns out to involve
less trivial renormalization conditions.

\subsection{Sine-Gordon kink}

The Lagrangian of the sine-Gordon model is 
\be
{\cal L}= -\frac{1}{2}\partial_\mu\varphi\partial^\mu\varphi+\frac{m^4}{\lambda}\[\cos\left(\frac{\sqrt{\lambda}}{m}\varphi\right)-1\] \, ,
\ee
with a real scalar field $\varphi$. We have a discrete $\mathbb{Z}_2 \times \mathbb{Z}$ symmetry, 
given by $\varphi\to -\varphi$ and \mbox{$\ph\to\ph+2\pi\, n {m\0\sqrt{\lambda}}$} with $n\in\mathbb{Z}$. 
The field equation in 1+1 dimensions is
\be \label{eomSG}
\partial_\mu\partial^\mu\varphi\equiv
(-\6_t^2+\6_x^2)\,\varphi=\frac{m^3}{\sqrt{\lambda}}\sin\left(\frac{\sqrt{\lambda}}{m}\varphi\right).
\ee
The constant solution of this equation yields the classical vacua 
\be
\ph=2\pi n{m\0\sqrt{\lambda}} \, ,
\ee
which break the discrete symmetry spontaneously. 
The kink solution interpolating between
the vacua with $n=0$ ($x=-\infty$) and $n=1$ ($x=+\infty$) is 
\be \label{kink0SG}
\ph_K(x) = \frac{4m}{\sqrt{\lambda}}\arctan e^{m(x-x_0)} \, .
\ee
The energy of this configuration is $M=8m^3/\lambda$. The equation for the fluctuations $\eta(x,t)$ around the kink solution,
\be
\partial_\mu\partial^\mu\eta(x,t)-  m^2\eta(x,t)\cos\left[\frac{\sqrt{\lambda}}{m}\ph_K(x)\right] =0 \, ,
\ee
becomes, using $\eta(x,t)=e^{-i\omega t}\phi(x)$ and the kink profile (\ref{kink0SG}), 
\be\label{diffeqSG}
\left[-\partial_x^2+m^2\left(1-\frac{2}{\cosh^2mx}\right)\right]\phi(x) = \omega^2\phi(x) \, ,
\ee
where we have set $x_0=0$. From this equation we obtain the zero-mode with energy $\omega =0$, 
\be\label{phi0SG}
\phi_0(x) = \sqrt{\frac{m}{2}}\frac{1}{\cosh mx} \, ,
\ee
and the continuum with $\omega_k=\sqrt{k^2+m^2}$,
\be
\phi_k(x) = \frac{m}{\omega_k}e^{ikx}\left(\tanh mx -i\frac{k}{m}\right) \, .
\ee
We note the useful relation between the continuum and the zero-mode
(Ref.~\cite{Goldhaber:2001rp}, Eq.~(12))
\be \label{useful}
|\phi_k(x)|^2=1-\frac{2m}{\omega_k^2}\phi_0^2 \, .
\ee

We can now compute the correction to the kink profile in the presence of the fluctuations. To this end, following Ref.\ \cite{Goldhaber:2001rp},
we interpret the equation of motion (\ref{eomSG}) as a Heisenberg field equation and write $\ph(x,t)\to \ph_K(x) + \ph(x,t) =\ph_{K}(x)+\phi_1(x)+\eta(x,t)$. Here, $\phi_1(x)\equiv \<\ph(x,t)\>$, and the 
quantum fluctuation field $\eta(x,t)\equiv \ph(x,t)-\phi_1(x)$ obeys $\<\eta\>=0$. 

The equation of motion for $\phi_1(x)$ becomes
\be \label{diffeq3}
\left[\partial_x^2-m^2\cos\left(\frac{\sqrt{\lambda}}{m}\ph_K\right)\right]\phi_1=-\frac{m\sqrt{\lambda}}{2}
\sin\left(\frac{\sqrt{\lambda}}{m}\ph_K\right)\langle\eta^2\rangle_{\rm ren} \, ,
\ee
where
\be \label{etarenSG}
\langle\eta^2\rangle_{\rm ren} = \langle\eta^2\rangle - \left. \langle\eta^2\rangle\right|_{x\to \infty} 
\ee
is the renormalized propagator. 
To obtain Eq.\ (\ref{diffeq3}) from Eq.\ (\ref{eomSG}) we have 
employed the following renormalization.
We have fixed $\delta m^2$ and $\delta\lambda$ in the renormalization of the mass $m^2\to m_0^2=m^2+\delta m^2$ and the coupling constant 
$\lambda\to \lambda + \delta\lambda$ such that all one-loop (finite-temperature) graphs with one vertex cancel \cite{Coleman:1975bu}. This condition 
implies that the ratio $m^2/\lambda$ is unchanged under renormalization, and 
\be\label{dvSG}
\delta m^2 = m^2\,\frac{\delta\lambda}{\lambda}=
\frac{\lambda}{2}\left. \langle\eta^2\rangle\right|_{x\to \infty} \, .
\ee

The propagator at finite temperature is given by
\be\label{eta2unren}
\langle\eta^2\rangle = T\sum_n\int_{-\infty}^{\infty} \frac{dk}{2\pi}\frac{|\phi_k(x)|^2}{\omega_n^2+\omega_k^2} 
= \int_{-\infty}^{\infty} \frac{dk}{2\pi}\frac{1+2n(\omega_k)}{2\omega_k}|\phi_k(x)|^2
\, ,
\ee
where $\omega_n = 2n\pi T$ are the bosonic Matsubara frequencies
and $n(\omega)=[\exp(\omega/T)-1]^{-1}$ is the Bose-Einstein distribution.
Here we have dropped the zero mode of Eq.~(\ref{phi0SG}), which corresponds to
the translational degree of freedom of the kink and which can be
taken care of by collective quantization \cite{Gervais:1975dc}.
This expression is UV divergent and thus requires regularization. In Sect.~\ref{sectph}
we shall introduce our method of dimensional regularization adapted to solitons, but in this and the next section we shall suppress its details, since the finite results
for the kink profile below do not depend on them. (If we were to calculate also
local energy densities, as we shall do for the model of Sect.~\ref{sectph}, we
would need to be more careful.) The treatment of the zero mode in
our way of dimensional regularization will also be discussed more fully
in Sect.~\ref{sectph}, where we cover the more general case of domain walls.

Simply subtracting Eq.\ (\ref{eta2unren})
according to Eq.\ (\ref{etarenSG}), and making use of 
Eq.\ (\ref{useful}) yields the renormalized propagator
\be \label{etarenSG1}
\langle\eta^2\rangle_{\rm ren}= -\phi_0^2(x)\left[\frac{1}{m\pi}+2m\int_{-\infty}^{\infty}\frac{dk}{2\pi}\frac{n(\omega_k)}{\omega_k^3}\right] \, .
\ee
Inserting this result into the differential equation (\ref{diffeq3}) yields the correction to the kink profile
\be \label{kink2}
\phi_1(x)=\frac{\sqrt{\lambda}}{4m}\left(\frac{1}{2\pi}+m^2\int_{-\infty}^\infty
\frac{dk}{2\pi}\frac{n(\omega_k)}{\omega_k^3}\right)\frac{\sinh mx}{\cosh^2 mx} \, .
\ee
The total kink is then given by $\varphi_K(x)+\phi_1(x)$ with $\varphi_K$ from Eq.\ (\ref{kink0SG}). We see that there is a zero-temperature
and a finite temperature correction, both being proportional to the derivative
of the zero mode. This is a consequence of the
renormalization (\ref{dvSG}) which 
subtracts the complete one-loop (seagull) diagram of the trivial sector
including thermal contributions. Had we left out the latter (or any
finite part), this would have produced extra contributions
to $\phi_1$ proportional to $m{\6\0\6 m}\varphi_K=\varphi_K+{2m\0\sqrt{\lambda}}\,mx/\cosh(mx)$.

\begin{figure}[tb]
\begin{center}
\includegraphics[width=0.5\textwidth]{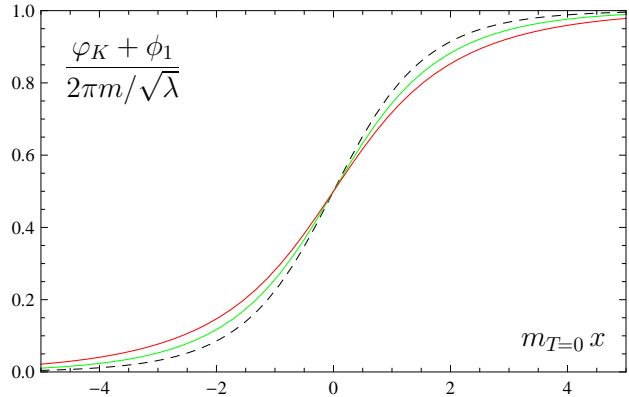} 
\caption{Finite-temperature corrections to the sine-Gordon kink $\varphi_K(x)+\phi_1(x)$ according to the result (\ref{kink2}) for
$\lambda/m^2 =0.2$ as a function of $x$ times the zero-temperature mass $m_{T\!=0}$
and for three different temperatures, $T/m_{T\!=0}=0,10,20$. The zero-temperature
result is given by the dashed line, and with increasing temperature the kink
becomes flatter.} 
\label{figkinkSG}
\end{center}
\end{figure}

With the 
large-temperature expansion 
\be
m^2\int_{-\infty}^\infty\frac{dk}{2\pi}\frac{n(\omega_k)}{\omega_k^3} = \frac{T}{2m}-\frac{1}{4\pi}+{\cal O}\left(\frac{m}{T}\right) \, ,
\ee
we see that the finite-temperature part of $\phi_1$ is suppressed compared to $\varphi_K$ by one power of the expansion parameter
$\lambda T/m^3$ (while the zero-temperature part is suppressed by $\lambda/m^2$). Since this expansion parameter cannot be small in the 
case of symmetry restoration (see discussion in the introduction), this perturbative result is only valid for temperatures much 
smaller than the critical temperature of symmetry restoration.

Identifying zero-temperature and thermal contributions of the
mass counterterm $\delta m^2\equiv \delta_{T=0} m^2+\delta_T m^2$, we
note that our renormalized mass $m^2\equiv
m_0^2-\delta m^2=(m_0^2-\delta_{T=0} m^2)-
\delta_T m^2=m^2_{T=0}+m^2_T$ differs from the renormalized mass at zero temperature
by a negative thermal correction
\be\label{m2TsG}
m^2_T=-\delta_T m^2=-{\lambda\02}\int_{-\infty}^\infty \frac{dk}{2\pi}\frac{n(\omega_k)}{\omega_k}=-{\lambda\04}{T\0m}+\ldots\quad\mbox{for}\;T\gg m.
\ee
This means that thermal corrections tend to flatten out the potential
by reducing the difference between maxima and minima of the potential, which
is proportional to $m^2(m^2/\lambda)$ with $m^2/\lambda$ invariant.
However, the distance between the minima and thus
the value of $\varphi_K$ at $x=\pm\infty$ remains fixed.
In Fig.\ \ref{figkinkSG} we plot the kink profile for three different 
temperatures as a function of $\,m_{T\!=0}x$, showing that with increasing
temperature the kink profile becomes flatter. (Had we plotted 
the profile as a function
of $mx$ with $m$ the temperature-dependent mass, the kink would have
appeared to become steeper instead.)

\subsection{CP$^1$ kink}

The discussion of the finite-temperature kink in the mass-deformed CP$^1$ model is very similar to the sine-Gordon model of the previous
subsection. The Lagrangian of the massive CP$^1$ model is
\be\label{L3d}
\mathcal L=-{r\0 (1+\phi^\dagger\phi)^2}\left(
\6_\mu \phi^\dagger \6^\mu\phi + m^2\phi^\dagger\phi \right)\,  ,
\ee
with a dimensionless coupling $r$ and a complex field $\phi$. 
This model is renormalizable in 1+1 dimensions and requires
a coupling constant
counterterm $r\to r+\delta r$ which is equivalent to
wave function renormalization
and a mass counterterm $m^2\to m^2+\delta m^2$.

The Lagrangian (\ref{L3d})
has a $U(1)$ symmetry, $\phi\to e^{i\alpha}\phi$, and a discrete
$\mathbb{Z}_2$ symmetry, $\phi\to 1/\phi^\dag$. Since we work in 1+1 dimensions, the continuous symmetry cannot be spontaneously broken 
\cite{Coleman:1973ci}. A vacuum expectation value of the field $\phi$
rather breaks the discrete 
$\mathbb{Z}_2$ symmetry spontaneously. The classical potential 
\be\label{VCP1}
V=\frac{rm^2\phi^\dagger\phi}{(1+\phi^\dagger\phi)^2} 
\ee
is minimized at $\phi=0$ and $|\phi|=\infty$, i.e., by the south and north poles of the 2-sphere 
representing the compactified complex plane. (Both minima are invariant under $U(1)$ as required). 

For a static solution, we can rewrite the Hamiltonian as
\be
\mathcal H = {r\0(1+\phi^\dagger\phi)^2}
\( \6_x\phi^\dagger-m\phi^\dagger\)\(\6_x\phi-m\phi\)
-rm\6_x\(1+\phi^\dagger\phi\)^{-1}
\ee
and read off the Bogomolnyi equation $(\6_x-m)\phi=0$.
Thus the classical solution is
\be \label{phiK}
\ph_K(x)=e^{m(x-x_0)} \, ,
\ee
and the classical energy is $M=rm$. 
We again set $x_0=0$.

Next we determine the spectrum of the 
fluctuations in the presence of the kink.
From
\be\label{eqA}
{\delta \mathcal L\0\delta\phi^\dagger}={r\0\rho^3}\left[\rho
(\Box-m^2)\phi-2\phi^\dagger((\6_\mu\phi)^2-m^2\phi^2)\right],\quad
\rho \equiv 1+\phi^\dagger\phi,
\ee
it is clear that the field equation of the fluctuations is obtained
by setting $\phi=\eta$, keeping $\rho$ fixed. This yields
\be \label{fluct}
\left(\partial_0^2-\partial_x^2+\frac{4\ph^2_K}{1+\ph_K^2}m\partial_x+m^2\frac{1-3\ph_K^2}{1+\ph_K^2}\right)\eta(x,t)=0
\, .
\ee
With the ansatz
\be
\eta(x,t)=r^{-1/2}(1+\ph_K^2)\,g(x)e^{-i\omega t}
\ee
the field equation for $g(x)$ becomes Eq.\ (\ref{diffeqSG}) of the sine-Gordon model.
Consequently, as in the sine-Gordon model,
we have a zero-mode
\be\label{g0x}
g_0(x) = \sqrt{\frac{m}{2}}\frac{1}{\cosh mx} \, ,
\ee
and the continuous spectrum $\omega_k=\sqrt{k^2+m^2}$ with 
\be
g_k(x) = \frac{m}{\omega_k}e^{ikx}\left(\tanh mx -i\frac{k}{m}\right) \, .
\ee

To obtain the corrections to the kink profile induced by the fluctuations we write, analogous to above, 
$\phi(x,t)=\ph_K(x)+\phi_1(x) +\eta(x,t)$, 
and derive the equation for $\phi_1(x)$.
The terms with  $\<\eta^* \eta\>$ are obtained by expanding the term
$-2(\phi^*/\rho)[2\6_x\ph_K\6_x\eta-2m^2\ph_K\eta]$ 
in Eq.~(\ref{eqA}) to first order
in $\eta^*$.
It is obvious from (\ref{eqA}) that the propagator is diagonal: $\<\eta\eta\>=0$.
Hence we may set $\phi=\varphi_K+\phi_1(x) +\eta_1(x,t)$ with real
$\phi_1(x)$ and $\eta_1(x)$ and replace $\<\eta^* \eta\>$ by
$\<\eta_1\eta_1\>$. Taking into account also the mass counterterm\footnote{%
The coupling constant counterterm does not contribute because
it amounts to wave function renormalization and thus multiplies
the classical field equation.}
this yields
\bea \label{diffeq4}
\left(\partial_x^2-\frac{4\ph_K^2}{1+\ph_K^2}m\partial_x-\frac{1-3\ph_K^2}{1+\ph_K^2}m^2\right)\, \phi_1
=\frac{4\ph_K m}{(1+\ph_K^2)^2}\langle\eta_1(\partial_x-m)\eta_1\rangle
+\delta m^2 \ph_K \frac{1-\ph_K^2}{1+\ph_K^2}
\, ,
\eea
with 
\be
\langle \eta_1^2\rangle = 
r^{-1}(1+\ph_K^2)^2 
\int_{-\infty}^\infty \frac{dk}{2\pi} 
\frac{1+2n(\omega_k)}{2\omega_k}|g_k(x)|^2
\, .
\ee
Using that $\langle\eta_1(\partial_x-m)\eta_1\rangle = \left(\frac{1}{2}\partial_x-m\right)\langle\eta_1^2\rangle$
and $\left(\frac{1}{2}\partial_x-m\right)(1+\ph_K^2)^2=
m(\ph_K^2-1)(1+\ph_K^2)$
we find that a mass counterterm of the form
\be\label{dm2CP1}
\delta m^2=4 r^{-1}
m^2 \int_{-\infty}^\infty \frac{dk}{2\pi} \frac{1+2n(\omega_k)}{2\omega_k}
\ee
corresponds to a UV-subtracted quantity
\bea
\langle\eta_1^2\rangle_{\rm ren} 
&=& r^{-1}(1+\ph_K^2)^2 \int_{-\infty}^\infty \frac{dk}{2\pi} \frac{1+2n(\omega_k)}{2\omega_k}\left[|g_k(x)|^2-1\right] \nn
&=& -
r^{-1}(1+\ph_K^2)^2g_0^2(x)\left(\frac{1}{m\pi}+2m \int\frac{dk}{2\pi}\frac{n(\omega_k)}{\omega_k^3}\right)
 \, ,
\eea
since
the functions $g_k(x)$ and $g_0(x)$ obey the same identity (\ref{useful})
as in the sine-Gordon case. With the explicit form of $g_0(x)$, 
Eq.~(\ref{g0x}), 
we finally obtain a vanishing result for the 
right-hand side of the differential equation (\ref{diffeq4}),
\be
\langle\eta_1(\partial_x-m)\eta_1\rangle_{\rm ren} = \left(\frac{1}{2}\partial_x-m\right)\langle\eta_1^2\rangle_{\rm ren}=0 \, ,
\ee
leading to the remarkably simple result 
\be\phi_1(x)=0.\ee 
This means that the kink profile remains unaltered at finite temperature, provided the mass
is renormalized according to (\ref{dm2CP1}).
The corresponding renormalization condition turns out to be an on-shell
mass renormalization in the trivial sector
including all thermal contributions\footnote{In the
CP$^1$ model the one-loop
self-energy diagram is momentum-dependent, but it remains Lorentz-invariant
at finite temperature, so that one does not have to distinguish between
e.g.\ a screening mass or a plasmon mass.}.
The thermal correction to the mass has the same form as in
the sine-Gordon model, with the replacement
$\lambda\to 8r^{-1}m^2$ in Eq.~(\ref{m2TsG}).
Since it is negative, this means that the kink $e^{mx}$ interpolating
between the minima at $\phi=0$ and $\phi=\infty$ becomes spread out
with increasing temperature.
For the height of the potential between these minima, also the
coupling constant renormalization is required. Direct calculation shows that
the one-loop self-energy equals $(p^2+3m^2)$ times a
momentum independent expression, which
implies that wave-function renormalization gives
the counterterm $\delta r/r=\2 \delta m^2/m^2$. Hence, the thermal
corrections of
mass and wave function both work in the direction of diminishing
the potential given in Eq.~(\ref{VCP1}) as the temperature is increased.

Once again we can only determine the onset of the melting of the
kink. The above discussion for the sine-Gordon model applies
with $r^{-1}$ replacing the dimensionless coupling parameter $\lambda/m^2$.
The requirement of a small 
expansion parameter at finite temperature, $r^{-1}T/m\ll1$,
precludes the consideration of
symmetry restoration as this would need $\delta m_T^2\gtrsim m^2$.

\section{$\ph^4$ kink and domain walls}\label{sectph}

The Lagrangian for the kink with $\lambda\ph^4$ interaction is
\bel{Lphi4}
\CL=-\2 \6_\mu \ph \6^\mu \ph -{\lambda\04}(\ph^2-v^2)^2\, ,
\quad v^2\equiv \mu^2/\lambda\, ,
\ee
with $\mathbb{Z}_2$ symmetry $\ph\to-\ph$.

To one-loop order we 
introduce counterterms $v^2\to v^2+\delta v^2$ and $\lambda\to\lambda+
\delta\lambda$ and $\delta \mu^2\equiv \delta(\lambda v^2)=\lambda\delta v^2+v^2\delta \lambda$
(leaving out wave-function renormalization which
would be needed only at two-loop order in 3+1 dimensions). $\delta v^2$
will be chosen such that in the topologically trivial sector 
tadpole contributions
are subtracted completely (including thermal contributions). 
In the absence of
wave-function renormalization, 
$\delta\lambda$ is fixed by a renormalization
condition for the mass of fluctuations in the trivial sector,
which at tree level is $m^2=2\mu^2$. (Note however that in the
spontaneously broken model (\ref{Lphi4}) the
counterterms $\delta v^2$ and $\delta\lambda$ imply the mass
counterterm
$\delta m^2=-\lambda\delta v^2+2 v^2\delta\lambda
\not\equiv 2\delta\mu^2$.) At zero temperature, the various possibilities have
been discussed in detail in Ref.~\cite{Rebhan:2002uk}. 
Below we shall address this
question at finite temperature and single out one particularly natural
renormalization condition.

The equation of motion in the $\varphi^4$ model reads
\bel{eomphi4}
\6_\mu \6^\mu \ph + \mu^2 \ph - \lambda \ph^3 = 0.
\ee
A kink at rest at $x=x_0$ 
which interpolates between the two degenerate vacuum states
\be
\ph=\pm \frac{\mu}{\sqrt{\lambda}} \equiv \pm v
\ee
is classically
given by \cite{Raj:Sol}
\bel{Ksol}
\ph_{K}(x)=v 
\tanh\(\mu(x-x_0)/\sqrt2\),
\ee
and its energy at tree-level is $M=(\sqrt2\mu)^3/3\lambda$. From 
now on we set $x_0=0$ without loss of generality.

The kink can be trivially embedded in $d$+1 spacetime dimensions where it
represents a domain wall separating the two vacua of the model.
In this case, $M$ has the meaning of a surface tension, i.e.\
energy per unit transverse volume. In the following we shall
make $d$ continuous and use this for dimensional regularization
in our renormalization program \cite{Bollini:1984jm,Parnachev:2000fz,Rebhan:2002uk}.

Fluctuations $\eta(\vec x,t)$ 
about the classical kink solution
are simple plane waves in the transverse directions $\vec y$
with transverse momentum $\vec\ell$, i.e., $\eta(\vec x,t)=
\int\!\!\!\!\!\Sigma
e^{-i\omega_{i\ell} t+i\vec\ell\cdot\vec y}\phi_i(x)$, with
$\omega_{i\ell}^2\equiv \omega_i^2+\ell^2$,
where the $x$-dependent
part 
$\phi_i(x)$ is then given by the 1+1-dimensional fluctuation equation
\be
\(-\partial_x^2-\mu^2+3\lambda\phi_K^2\)\phi_i(x)=\omega_i^2\phi_i(x).
\ee

The spectrum of the 1+1-dimensional fluctuation equation \cite{Raj:Sol}
has a zero-energy solution ($\omega_0=0$),
\be
\phi_0(x)=\sqrt{\frac{3m}{8}}\,\frac{1}{\cosh^2\frac{mx}{2}} \, ,
\ee
a bound state with energy $\omega_B=\sqrt{3}\,m/2$,
\be
\phi_B(x)=\sqrt{\frac{3m}{4}}\,\frac{\sinh\frac{mx}{{2}}}{\cosh^2\frac{mx}{2}} \, ,
\ee
and a continuous spectrum with energies $\omega_k=\sqrt{k^2+m^2}$,
\be
\phi_k(x) = \frac{m^2e^{ikx}}{4\omega_k\sqrt{\omega_k^2-\omega_B^2}}\left[-3\tanh^2\frac{mx}{{2}}+1+4\frac{k^2}{m^2}+6i\frac{k}{m}
\tanh\frac{mx}{{2}}\right] \, .
\ee
Below we shall use the relation \cite{Goldhaber:2001rp}
\be \label{help}
|\phi_k(x)|^2=1-\frac{2m}{\omega_k^2}\phi_0^2(x)-\frac{m}{\omega_k^2-\omega_B^2}\phi_B^2(x) \, .
\ee

\subsection{Kink profile}

Interpreting (\ref{eomphi4}) as a Heisenberg field equation
and writing $\ph(x,t)=\ph_{K}(x)+\phi_1(x)+\eta(x,t)$ with the
quantum fluctuation field obeying $\<\eta\>=0$,
we have the following equation
for the one-loop correction to the kink profile,
\beal{phi1K}
\[\6_x^2+\mu^2-3\lambda\ph_K^2(x)\]\phi_1(x)&=&
3\lambda\ph_K(x)\[\<\eta^2\>(x)
-{1\03}\delta v^2-{\delta\lambda\03\lambda}(v^2-\ph_K^2(x))\]\nn
&\equiv& 3\lambda\ph_K(x)\;\<\eta^2\>_{\rm ren.}(x)
\,.
\eea
This equation is formally 
unchanged in the presence of a nonzero wave-function renormalization $Z$, 
although
the counterterms $\delta v^2$ and $\delta\lambda$ will of course
be different in different renormalization schemes.
However, if two renormalization schemes, one with $Z=1$ and
one with $Z\not=1$, have the same renormalization
condition for the mass $m$ (e.g. that it should be
the pole mass in the trivial sector without a kink), 
the one with nontrivial $Z$ differs
from the other one only by an extra contribution\footnote{As one
can check (cf.\ Sect.~2.2.2 of Ref.~\cite{Rebhan:2002uk}), the net
difference of the r.h.s. of (\ref{phi1K}) is then $+\lambda(Z-1)\ph_K^3$,
which leads to the extra contribution $\Delta\phi_1=-\2(Z-1) \ph_K$.
}
$\Delta\phi_1=-\2(Z-1) \ph_K$.

\begin{figure}
\includegraphics[bb=150 535 373 780,clip,scale=0.8]{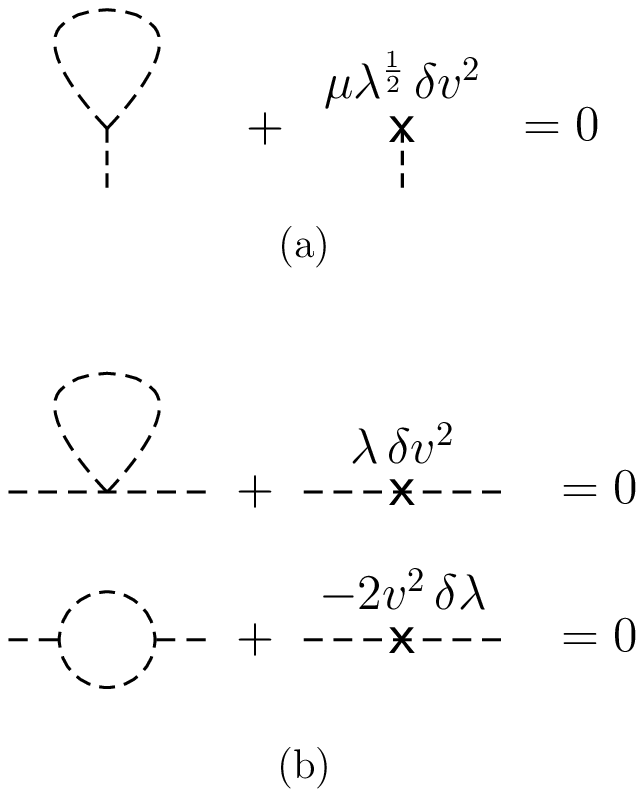}
\caption{Renormalization of the tadpole diagram (a) through the
counterterm $\delta v^2$ and of the self-energy diagrams (b)
through the counterterm $-\delta m^2=\lambda\delta v^2-2v^2\delta\lambda$.
Dashed lines correspond to the propagator in the trivial sector.
\label{figct}}
\end{figure}

Continuing with $Z=1$, we have
\be\label{dv2th}
\delta v^2=3 \left.\<\eta^2\>\right|_{\rm trivial\; sector}=
3\int {d^{d-1}\ell\,dk\0(2\pi)^d} {1+2n(\wkl)\02\wkl}
\ee
where we absorb the complete quantum and thermal contribution of
the tadpole diagram in the renormalization of $v$ (see Fig.~\ref{figct}a).
Writing everything out for $d$+1 dimensions, this leaves us with
\beal{eta2ren}
\<\eta^2\>_{\rm ren.}(x)&=&
\int {d^{d-1}\ell\,dk\0(2\pi)^d} {1+2n(\wkl)\02\wkl}
\biggl[-{3m^2/4\0k^2+{m^2\04}}{1\0\cosh^2(mx/2)}\nn
&&\hspace{45mm}
+{(3m^2/4)^2\0(k^2+m^2)(k^2+{m^2\04})}{1\0\cosh^4(mx/2)}
\biggr]\nn
&&+\int {d^{d-1}\ell\0(2\pi)^{d-1}} {1+2n(\wbl)\02\wbl}
{3m\04}\({1\0\cosh^2(mx/2)}-{1\0\cosh^4(mx/2)}\)\nn
&&+\mathcal N_d(T) {3m\08}{1\0\cosh^4(mx/2)}-\delta\lambda{m^2\06\lambda^2}{1\0\cosh^2(mx/2)},
\eea
where $\wkl=\sqrt{k^2+\ell^2+m^2}$ and $\wbl=\sqrt{\ell^2+3m^2/4}$. 
We have now also included the zero mode of the 1+1-dimensional fluctuation
equation about the kink background, which in $d>1$ spatial dimensions
is a massless mode (the Goldstone mode associated with the
spontaneous breaking of translational invariance),
\be\label{Nd}
\mathcal N_{d>1}(T)=\int {d^{d-1}\ell\0(2\pi)^{d-1}} {1+2n(\wol)\02\wol},
\quad \wol=\sqrt{\ell^2},
\ee
where the $T=0$ part vanishes since in dimensional regularization
scaleless integrals are identically zero.
For $d=1+\epsilon$ we discard the thermal part as well,
since the latter is finite and we can set $\epsilon=0$ and $\wol=0$ there,
assuming that this genuine zero mode is treated by collective quantization
\cite{Gervais:1975dc}. Thus we have 
\be\mathcal N_{d=1+\epsilon}(T)\equiv 0;\quad \mathcal N_d(T=0)=0.
\ee
In keeping a nontrivial thermal contribution $\mathcal N_d$ for $d\ge2$
we differ from Ref.~\cite{Graham:2001kz}, and we shall later in Sect.~\ref{sec:led}
provide evidence for the necessity of this.

As mentioned above, in $d=3$, $\delta\lambda$ which appears in the
last term of Eq.~(\ref{eta2ren}) has to absorb a UV divergence
of the momentum integrals, whereas in lower dimensions one could
also be content with a minimal renormalization scheme, where only
tadpoles are renormalized.
In Refs.~\cite{Shifman:1998zy,Goldhaber:2001rp} 
it was observed (for $d=1$) that
at zero temperature the requirement of an on-shell mass simplifies
the result greatly -- all terms proportional to $1/\cosh^2(mx/2)$ then cancel.

At nonzero temperature, there is actually not just one on-shell mass,
but thermal masses are in general momentum dependent: at zero momentum,
the thermal mass gives the plasma frequency above which
propagating modes appear. Because of the absence of manifest
Lorentz invariance, the effective mass of the propagating mode is 
generally not a constant.
Thermal masses at frequencies below the plasma frequency give
inverse spatial screening lengths. As we shall show now, it is
the static screening mass which has to be employed in the
renormalization condition in order that the above mentioned
simplification occurs.

Requiring that the renormalized parameter $m$ be equal to
the thermal static screening mass means that we have to
subtract
the self-energy diagram at zero frequency and imaginary spatial 
momentum\footnote{In
a nonabelian gauge theory, one finds that only this self-consistent
definition leads to a gauge-independent result for
the (Debye) screening mass \cite{Rebhan:1993az}. In the present
scalar theory, it will be seen to remove certain artefacts in the
kink profile.}
$\vec q^{\,2}=-m^2$. 
The mass counterterm follows from (\ref{Lphi4}) after substituting
$\ph=v+\eta$, and reads $\delta m^2=-\lambda\delta v^2+2 v^2\delta\lambda$.
As indicated in Fig.~\ref{figct}b, the seagull
diagram is already cancelled by the counterterm $-\lambda\delta v^2$, so the
diagram with two propagators in the loop evaluated at $\vec q^{\,2}=-m^2$
defines $2 v^2\delta\lambda$.
Using Feynman parametrization (which is straightforward to use
in finite-temperature integrals at zero frequency, albeit not
otherwise \cite{Weldon:1993bv}) we find 
\bel{dlscr}
\delta\lambdad=
{9\lambda^2\02} \int_0^1 dt
\int {d^{d-1}\ell\,dk\0(2\pi)^d} \[{1+2n(\omega_t)\02\omega_t^3}
-{n'(\omega_t)\0\omega_t^2}\],
\ee
where
\bel{omt}
\omega_t\equiv \sqrt{k^2+\ell^2+m^2[1-t(1-t)]}.
\ee
In the $T=0$ part of $\delta\lambdad$, one can easily integrate
over the Feynman parameter to find
\be
\delta\lambdad^{(T=0)}
={9\lambda^2\04}
\int {d^{d-1}\ell\,dk\0(2\pi)^d} {1\0\sqrt{k^2+\ell^2+m^2}(k^2+\ell^2+3m^2/4)}.
\ee
For later use we quote the closed-form result for this counterterm
from Ref.~\cite{Rebhan:2002uk},
\bel{deltalambda2F1}
\delta\lambdad^{(T=0)}
=9\lambda^2 {m^{d-3}\0(4\pi)^{d+1\02}}\textstyle{\Gamma({3-d\02})
\left(3\04\right)^{d-3\02} {}_2F_1({3-d\02},{1\02};{3\02};-{1\03})}.
\ee
However, keeping the integral over transverse momenta $\ell$ unevaluated
and using
\be
\int_{-\infty}^\infty dk{1\0\sqrt{k^2+a^2}(k^2+b^2)}=
\frac{2}{b \sqrt{a^2-b^2}}\arctan\left(\frac{\sqrt{a^2-b^2}}{b}\right)
\ee
one can verify that all zero-temperature 
terms proportional to $1/\cosh^2(mx/2)$
in (\ref{eta2ren}) cancel already under the integral over the transverse
momentum $\ell$. Thus the cancellations observed in Refs.~\cite{Shifman:1998zy,Goldhaber:2001rp} for zero temperature and $d=1$ also take place for
arbitrary $d$.

\begin{figure} [ht]
\begin{center}
\includegraphics[width=0.5\textwidth]{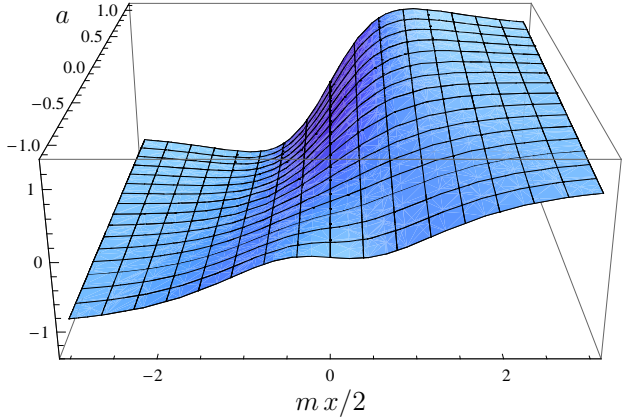}
\caption{The one-loop corrected
$\ph^4$ kink profile $\ph_K(x)+\phi_1(x)\propto \tanh(z)+a\sinh(z)/\cosh^3(z)$ with $z=mx/2$
for $a=-1\ldots1$, showing the effect of positive vs. negative $a\propto A_d$
up to nonperturbatively large $a$. 
For $d=1$, $A_d$ is positive and increasing with temperature,
as displayed in
Fig.~\ref{figa1}, seemingly leading to a steeper kink at higher
temperatures.
However, this plot does not yet take into account that at increasing
temperature the asymptotic values of the kink are reduced
by the thermal part of $\delta v^2$, see Eq.~(\ref{dv2th}). Including
this effect, the kink profile does become flatter.
} 
\label{figphi4kinkprofile}
\end{center}
\end{figure}

Turning to the thermal contributions one finds that the individual 
integrals involving the Bose-Einstein factor $n$ cannot be evaluated in
closed form. By numerical integrations one can however verify without
difficulty the rather abstruse identity 
\bea\label{RAI}
\int_0^1 dt \int {dk\02\pi} \[{n(\omega_t)\0\omega_t^3}
-{n'(\omega_t)\0\omega_t^2}\]&&\nn
+\int {dk\02\pi}{n(\sqrt{k^2+\ell^2+m^2})\0\sqrt{k^2+\ell^2+m^2}}
{1\0k^2+{m^2\04}}&=& \frac{1}{m}{n(\sqrt{\ell^2+{3\04}m^2})\0\sqrt{\ell^2+{3\04}m^2}}\;,
\eea
where $\omega_t$ was defined in Eq.~(\ref{omt}). Using this to
evaluate the Feynman parameter integral in $\delta\lambdad$,
Eq.~(\ref{dlscr}), and inserting the result into Eq.~(\ref{eta2ren})
shows that at finite temperature it is the renormalization prescription
of vanishing tadpoles and $m$ being the finite-temperature screening
mass which absorbs all terms proportional to $1/\cosh^2(mx/2)$
in (\ref{eta2ren}). We then have 
\bel{eta2renfinal}
\<\eta^2\>_{\rm ren.}(x)=-m^{d-1}A_d(T/m){1\0\cosh^4(mx/2)}
\ee
with
\beal{AdTm}
A_d(T/m)&=&{3m^{2-d}\04}\int {d^{d-1}\ell\0(2\pi)^{d-1}} {1+2n(\wbl)\02\wbl}
-{3m^{2-d}\08} \mathcal N_d(T) \nn
&&-{9m^{5-d}\016}
\int {d^{d-1}\ell\,dk\0(2\pi)^d} {1+2n(\wkl)\02\wkl}
{1\0(k^2+m^2)(k^2+{m^2\04})}
\eea
which when inserted into Eq.~(\ref{phi1K}) yields
\bel{phi1A}
\phi_1(x)=v^{-2}m^{d-1}A_d(T/m){1\0\cosh^2(mx/2)}\ph_K
=v^{-1}m^{d-1}A_d(T/m){\sinh(mx/2)\0\cosh^3(mx/2)}
\ee
for the correction to the kink profile. In Fig.~\ref{figphi4kinkprofile}
the effect of such a correction to the classical kink profile is shown
for (large) negative and positive coefficients.

In a generic renormalization scheme, where 
$\<\eta^2\>_{\rm ren.}(x)$ has either uncancelled $1/\cosh^2(mx/2)$ terms or 
uncancelled constant terms (by incomplete tadpole
subtraction), one would find additional terms $\propto mx/\cosh^2(mx/2)$
in the kink profile correction $\phi_1$ (aside from a different
function $A_d(T/m)$). 
A posteriori, 
the ``bare'' $mx$ figuring in the
latter term 
can be understood as an artefact of incomplete renormalization
since the particular on-shell 
renormalization scheme considered above
is evidently
able to absorb these terms such that all $mx$ appear only in exponentiated
form.

So the simple result (\ref{phi1A}) depends on a renormalization scheme
where $m$ is the thermal static screening mass. This 
differs from
the zero-temperature mass by the finite difference
\be
m^2-m^2_{T=0}= 3\lambda\int {d^{d-1}\ell\,dk\0(2\pi)^d} {n(\wkl)\0\wkl}
\left(1+{3m^2/2\0k^2+{m^2\04}}\right)
-{9\lambda m\02}\int {d^{d-1}\ell\0(2\pi)^{d-1}} {n(\wbl)\0\wbl},
\ee
where we have again used the identity (\ref{RAI}).

\subsubsection{1+1 dimensional kink}

\begin{figure} [ht]
\begin{center}
\includegraphics[width=0.5\textwidth]{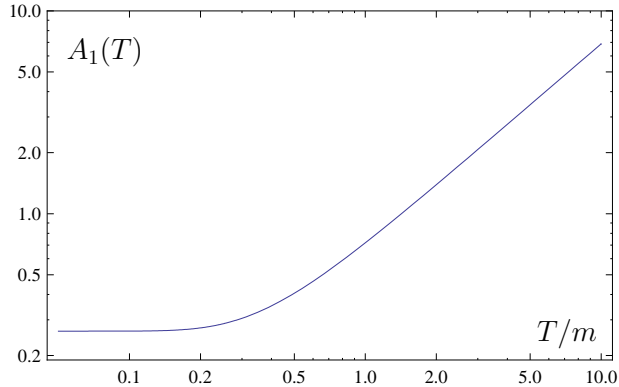}
\caption{The function $A_d(T/m)$ in a doubly logarithmic plot
for $d=1$. This function determines the temperature dependence 
of the kink profile, see
Eq.~(\ref{phi1A}).} 
\label{figa1}
\end{center}
\end{figure}

For $d=1$, $\mathcal N_d(T)$ is absent and all integrals in
(\ref{AdTm}) are individually finite 
and the $\ell$-integration can in fact be dropped. 
The $T=0$ part is readily found to be
\be\label{A10}
A_1(0)={1\04\sqrt3}+{3\08\pi},
\ee
in agreement with Refs.~\cite{Shifman:1998zy,Goldhaber:2001rp}.
The thermal part of $A_1(T)$ turns out to be strictly positive and
growing linearly with $T/m$ for large $T/m$. The full function is
plotted in Fig.~\ref{figa1}.
This corresponds to a positive parameter $a$ in Fig.~\ref{figphi4kinkprofile}
which grows as the temperature is increased (with $a$
having to remain sufficiently small so that
perturbation theory is still valid).
When plotted with fixed asymptotic values as in Fig.~\ref{figphi4kinkprofile},
the one-loop corrected
kink profile which is slightly steeper than the classical one
appears to become even steeper with increasing temperature.
However, the asymptotic amplitude of the kink diminishes,
because $v^2\equiv v_0^2-\delta v^2\equiv
(v_0^2-\delta_{T=0}v^2)-\delta_T v^2$, and $\delta_T v^2$,
the thermal part of Eq.~(\ref{dv2th}), is positive and growing with
temperature. Including this effect, the slope of the kink profile decreases 
at higher temperature. Moreover, also
$m$ has been renormalized such as to
include thermal corrections and these reduce $m$ as the phase transition
is approached. When plotted in terms of a fixed zero-temperature mass 
$m_{T=0}$, the kink profile flattens even more quickly.
At any rate, in perturbation theory we can only describe
reliably the onset of the melting of the kink.

The $\ph^4$ kink in 1+1 dimensions has been studied extensively in 
nonperturbative self-consistent
approximations to the two-particle irreducible effective action
\cite{Boyanovsky:1998ka,Salle:2003ju,Bergner:2003au}. 
Comparing our Fig.~\ref{figphi4kinkprofile}
with Figs.~2 and 3 
of \cite{Bergner:2003au} we find a somewhat different behavior
even when the corrections are still small.
This can be traced to the fact that 
in
Refs.~\cite{Boyanovsky:1998ka,Bergner:2003au} a variational ansatz
for the dressed two-point function with coinciding
arguments in the kink background, a nonperturbative generalization of our
$\<\eta^2\>_{\rm ren.}(x)$, has been employed which is
proportional to $1/\cosh^2(mx/2)$.
Our result (\ref{eta2renfinal}) however suggests that a more adequate
ansatz would involve $1/\cosh^4(mx/2)$. In Ref.~\cite{Boyanovsky:1998ka}
the choice of the former was motivated by an analysis of the sine-Gordon model,
where we have indeed obtained a renormalized two-point function
proportional to $1/\cosh^2(mx)$, see Eq.\ (\ref{etarenSG1}). But, as we have demonstrated,
the generalization to the $\ph^4$ model is not justified. In this case,
both the constant term in $\<\eta^2\>$ as well
as the term proportional to $1/\cosh^2(mx/2)$ are removable by
renormalization, whereas in the sine-Gordon model only the constant
term is. Note that in both cases the on-shell renormalized $\<\eta^2\>_{\rm ren.}$
to one-loop order
turns out to be proportional to the zero-mode of that model squared.

\subsubsection{Domain wall profiles}

For $d\ge2$, the zero-temperature contributions in Eq.~(\ref{AdTm})
are individually UV-divergent.
Carrying out first the $\ell$-integral in dimensional regularization we obtain
(using formula (3.259.3) of Ref.~\cite{GraR:T})
\beal{Ad0}
A_d(0)&=&{9\016}{1\0(4\pi)^{d/2}}\,\Gamma\({2-d\02}\)\[
\({3\04}\)^{{d-4\02}}\!-{1\0\pi}\int_0^\infty dx {(x^2+1)^{{d-4\02}}\0x^2+{1\04}}
\]\nn
&=&{9\016}{1\0(4\pi)^{d/2}}\,\Gamma\({2-d\02}\)\[
\({3\04}\)^{{d-4\02}}\!-{1\0\sqrt\pi}{\Gamma({5-d\02})\0\Gamma({6-d\02})}\;
{}_2F_1({4-d\02},{1\02};{6-d\02};{3\04})
\]\,.
\eea
In the limit $d\to 1$ this of course reproduces Eq.~(\ref{A10}),
but for $d\to 2$ we encounter a singularity, because the
square bracket in Eq.~(\ref{Ad0}) does not vanish
for $d\to2$. However, this divergence should not be an UV divergence,
since, as we have
shown above, mass, coupling constant as well as wave function
renormalization can only modify constant terms and terms
proportional to $1/\cosh^2(mx/2)$ in $\<\eta^2\>$ and therefore not $A_d$.
Indeed, by taking into account $\mathcal N_{2}(0)$, which is zero in
dimensional regularization, but whose integral representation is
both UV and IR divergent, one finds that a small IR regulator mass
$\ell^2\to\ell^2+\nu^2$ in (\ref{Nd}) changes 
$(\ref{Ad0})$ to
\bea
A_d(0;\nu/m)&=&{9\016}{1\0(4\pi)^{d/2}}\,\Gamma\({2-d\02}\)\[
\({3\04}\)^{{d-4\02}}\!-{2\03}\({\nu^2\0m^2}\)^{{d-2\02}}\!
-{1\0\pi}\int_0^\infty dx {(x^2+1)^{{d-4\02}}\0x^2+{1\04}}\]\nn
&=&{9\016}{1\0(4\pi)^{d/2}}\,\Gamma\({2-d\02}\)\[{4\03}-{2\03}-{2\03}+O(d-2)\],
\eea
which is finite for $d\to2$ but now contains a term involving $\ln(\nu/m)$, 
\bel{A20}
A_2(0)\sim -{3\016\pi}\ln{m\0\nu},
\ee
so that the IR regulator cannot be
removed. (The finite-$T$ contributions in $\mathcal N_d(T)$ 
have even more severe
IR divergences.)
Such IR divergences have in fact been discussed in the literature
as being associated with the roughening of interfaces, leading to
a logarithmic sensitivity of the width of interfacial profiles to
the linear system size $L$. This phenomenon has been described in terms of the
so-called capillary wave model \cite{Buff:1965zz,Gelfand:1990,Muller:2004vv}, with the capillary
waves corresponding to the massless modes associated with
the kink zero mode \cite{Rudnick:1978zz}. In a field
theoretic treatment similar to ours, but with the
interpretation of the Euclidean action
of the $\ph^4$ model as the Landau-Ginzburg Hamiltonian
of statistical field theory, interface roughening has been
discussed using cutoff regularization in Ref.~\cite{Bessa:2004pu},
and recently
in $d=3+\epsilon$
dimensions in Ref.~\cite{Kopf:2008hr}.
Our results for the
one-loop profile, Eq.~(\ref{phi1A}) and (\ref{A20}), agree 
with Refs.~\cite{Bessa:2004pu} and \cite{Kopf:2008hr}, the latter
upon identifying $\nu\sim 1/L$. 
(Ref.~\cite{Kopf:2008hr} used a finite system with quadratic interface
as IR regularization and also determined the resulting
sublogarithmic contributions.) 

The results for the profile
given in Refs.~\cite{Rudnick:1978zz,Kopf:2008hr}
also involve IR-finite contributions of the form
$mx/\cosh^2(mx/2)$, which are due to the fact that there the
results are expressed in terms of a static correlation length
defined by a renormalization point $q^2=0$ instead of
$q^2=-m^2$. As we have shown, the latter definition, which
is required to give $m$ the meaning of an inverse exponential
screening length, quite
generally eliminates such contributions.

For $d=3$, i.e., in the 3+1 dimensional case, 
we find that the $T=0$ part is finite, yielding the (to our
knowledge new) result
\be
A_3(0)=-{\sqrt3\032\pi}\,.
\ee
Now the thermal contributions are logarithmically IR divergent, because
of $\mathcal N_{3}(T)$, which leads to
\be
A_3(T)\sim -{3\016\pi}{T\0m}\ln{m\0\nu},
\ee
for an IR momentum cutoff $\nu\sim 1/L$, which
indicates interfacial roughening effects also in the
context of domain walls of relativistic field theories at
finite temperature.

\subsection{Local energy density}\label{sec:led}

We now turn to the local energy density
$\epsilon(x)$, which, following Ref.~\cite{Goldhaber:2001rp}, we
decompose as 
\be
\epsilon(x)=\epsilon_{\rm Cas}(x)+\Delta\epsilon_{\rm Cas}(x)+\Delta\epsilon_{(\phi_1)}(x).
\ee
Here $\epsilon_{\rm Cas}(x)$ represents the local energy density
of a domain wall in $d$+1 dimensions (assuming dimensional
regularization) due to the sum over zero-point energies, whereas
the contributions $\Delta\epsilon_{\rm Cas}(x)$ and $\Delta\epsilon_{(\phi_1)}(x)$ are total derivatives which do not contribute to the integrated
total energy (or surface tension) and which
have been identified in
Ref.~\cite{Goldhaber:2001rp}.

Subtracting off the energy density of the topologically trivial
sector (including thermal contributions), the local energy density
from the sum over zero-point energies is given by
\beal{epsCas}
\epsilon_{\rm Cas}(x)&=&
\int {d^{d-1}\ell\0(2\pi)^{d-1}}{\wol\02}[1+2n(\wol)]\phi_0^2(x)+
\int {d^{d-1}\ell\0(2\pi)^{d-1}}{\wbl\02}[1+2n(\wbl)]\phi_B^2(x)\nn
&&-m\int {d^{d-1}\ell\,dk\0(2\pi)^d}{\wkl\02}[1+2n(\wkl)]
\[{2\phi_0^2(x)\0k^2+m^2}+{\phi_B^2(x)\0k^2+{m^2\04}}\]\nn
&&-{\lambda\02}\delta v^2[\ph_K^2(x)-v^2]+{\delta\lambdad\04}[\ph_K^2(x)-v^2]^2,
\eea
where in the second line Eq.~(\ref{help}) 
has been used again to rewrite $|\phi_k(x)|^2-1$
in terms of $\phi_0(x)$ and $\phi_B(x)$,
and
where we have included the contribution from the counterterms.
(Recall that 
$\wkl=\sqrt{k^2+\ell^2+m^2}$, $\wbl=\sqrt{\ell^2+3m^2/4}$, and
$\wol=\sqrt{\ell^2}$.)
The first term in Eq.~(\ref{epsCas}) is the contribution from the
massless modes corresponding to the zero mode of the 1+1-dimensional kink.
Its zero-temperature part is eliminated by dimensional regularization, but
its thermal contribution is that of black-body radiation in $d-1$ spatial
dimensions. It is to be omitted for $d=1$, but contributes nontrivially
in the case of domain walls. If it was not included (as
done e.g.\ in Ref.\ \cite{Graham:2001kz}), the total
one-loop surface tension $\int dx\epsilon_{\rm Cas}$ would
not vanish in the limit $m\to0$, i.e.\ at the second-order phase transition
where the kink has melted completely. To see this, note that
for fixed $\ell\gg k$, one has
IR singular limits $m\to0$ of $\int dk({2\0k^2+m^2}+{1\0k^2+m^2/4})$
in the second line of Eq.~(\ref{epsCas}) which cancel 
the explicit factor of $m$
there. This yields thermal contributions
which indeed compensate the $m\to0$ limits of the thermal contributions of
the bound state modes and the massless modes.

In contrast to the case of the kink profile we observe that in $\epsilon_{\rm Cas}$
the massless modes do not lead to IR problems in either 2+1 or 3+1 dimensions,
but we shall see that the IR singular kink profile plays a role in the
remaining two (total derivative) contributions to $\epsilon(x)$.
The total energy (or surface tension) $M$ is
already determined by the above (IR-safe) expression,
\be
M=\int dx\,\epsilon_{\rm Cas}(x).
\ee
(The integration over $x$ is evaluated readily using
$\int dx \phi_B^2(x)=\int dx \phi_0^2(x)=1$.)

For the local distribution of the energy, however, the total derivative
terms $\Delta\epsilon_{\rm Cas}(x)$ and $\Delta\epsilon_{(\phi_1)}(x)$
are relevant.
The first of these 
comes from a surface term
associated to a partial integration of the spatial gradients
in the kinetic energy \cite{Goldhaber:2001rp} and reads
\bea\label{DepsCas}
\Delta\epsilon_{\rm Cas}&=&{1\04}\6_x^2\<\eta^2\>(x)\nn
&=&{1\04}\6_x^2\left[\<\eta^2\>_{\rm ren.}(x)
+\delta\lambda{m^2\06\lambda^2}{1\0\cosh^2(mx/2)}
\right],
\eea
with $\<\eta^2\>_{\rm ren.}$ defined in Eq.\ (\ref{phi1K}).

The second contribution to the local energy density 
comes from the correction to the kink profile, $\phi_1$
that we considered above. This also does not contribute to the total
energy, because the classical kink corresponds to a stationary point of
the classical energy, but it gives a local modification of the
energy density according to
\be
\Delta\epsilon_{(\phi_1)}=\6_x\(\phi_1\6_x\ph_K\).
\ee
Inserting the results of Eq.~(\ref{eta2renfinal}) and Eq.~(\ref{phi1A}) into
these two expressions, we find that the terms with the function 
$A_d$, which appears in both, add with equal magnitude, yielding
\be\label{DepsCasphi1}
\Delta\epsilon_{\rm Cas}+\Delta\epsilon_{(\phi_1)}=
\2 m^{d+1} A_d(T) \left(-{4\0\cosh^4{mx\02}}+{5\0\cosh^6{mx\02}}\right)
-\delta\lambdad{m^3\018\lambda^2}\[\phi_0^2(x)-\phi_B^2(x)\].
\ee
This localized contribution with zero total energy is finite in 1+1 dimensions,
where it has been evaluated at zero temperature in Ref.~\cite{Goldhaber:2001rp},
but in 2+1 and 3+1 dimensions it inherits the IR divergences found in
the prefactor $A_d$ of the one-loop kink profile discussed in the previous
section. In 3+1 dimensions it is moreover UV divergent because
of the appearance of $\delta\lambdad$ in $\Delta\epsilon_{\rm Cas}$.

Let us now check
for UV divergences in Eq.~(\ref{epsCas}),
which was evaluated in
\cite{Shifman:1998zy,Goldhaber:2001rp}
for the 1+1-dimensional kink, but only in integrated form
for $d$+1-dimensional domain walls in Ref.~\cite{Rebhan:2002uk}.
We shall now show that in the 2+1-dimensional case also
the local energy density is rendered finite by on-shell renormalization
(defined in the trivial sector). However, we shall find
that in 3+1 dimensions there are uncancelled UV divergences
in $\epsilon_{\rm Cas}^{(T=0)}(x)$ 
which require additional composite operator renormalization.

Potential UV divergences can only come from the $T=0$ part (terms without $n$)
in Eq.~(\ref{epsCas}).
Using that $\ph_K^2(x)-v^2=-{2m\03\lambda}(\phi_B^2+2\phi_0^2)$ and
$[\ph_K^2(x)-v^2]^2={2m^3\03\lambda^2}\phi_0^2$
we obtain
\bea
\epsilon_{\rm Cas}^{(T=0)}(x)&=&\phi_B^2(x)\;\frac{1}{2}\int {d^{d-1}\ell\0(2\pi)^{d-1}}\left[
\wbl+
m\int {dk\02\pi}\(
{1\0\wkl}-{\wkl\0k^2+{m^2\04}}
\)
\right]\nn
&&+\phi_0^2(x)\;m\int {d^{d-1}\ell\,dk\0(2\pi)^d}\[
{1\0\wkl}-{\wkl\0k^2+m^2}+{3\08}{m^2\0\wkl(\wkl^2-{m^2\04})}
\].
\eea
Integration over $\ell$ yields after some rearrangements
\bea
\epsilon_{\rm Cas}^{(T=0)}(x)&=&\phi_B^2(x)
{\Gamma(-{d\02})\0(4\pi)^{d/2}}\2\biggl[
-({3\04}m^2)^{d/2}
+ m (1-d) \int {dk\02\pi} (k^2+m^2)^{d/2-1}\nn
&&\qquad\qquad\qquad
+ m \int {dk\02\pi} (k^2+m^2)^{d/2-1} {3m^2/4 \0 k^2+m^2/4}
\biggr]\nn
&&+\phi_0^2(x)\biggl[ {\Gamma(-{d\02})\0(4\pi)^{d/2}}m
(1-d) \int {dk\02\pi} (k^2+m^2)^{d/2-1}
+\delta\lambdad^{(T=0)}{m^3\06\lambda^2}
\biggr],
\eea
with $\delta\lambdad^{(T=0)}$ given in (\ref{deltalambda2F1}).
This expression is finite for $d\to1$, and includes an
``anomalous'' term proportional to $(1-d)\int dk (k^2+m^2)^{d/2-1}$,
which would be missed in a naive momentum cutoff regularization.
With it one recovers Eq.~(18) of Ref.~\cite{Goldhaber:2001rp}, except
for the last term, which is due to $\delta\lambdad$ and which is
finite by itself\footnote{The inclusion of this term corresponds
to the finite renormalizations $m\to\bar m$ and $\lambda\to\bar\lambda$
in the final result (28) of Ref.~\cite{Goldhaber:2001rp}.}. 
For $d\to2$ the result is also finite (the divergent $\Gamma(-{d\02})$ which
is present in all terms except $\delta\lambdad$ is cancelled
after combining the UV finite expressions and using that
$\int dk (k^2+m^2)^{d/2-1}$ is proportional to $1/\Gamma(1-d/2)$),
but for $d=3-2\epsilon$ one finds divergent terms $1/\epsilon$. 
All remaining finite integrals over $k$ can be evaluated separately for
$d=1,2,3$ by using the substitution $k/m=\tan t$, but
using the formulae given in Ref.~\cite{Rebhan:2002uk}, the following
comparatively compact result can be derived, 
\beal{epsCasd}
\epsilon_{\rm Cas}^{(T=0)}(x)&=&
{m^{d}\0(4\pi)^{d+1\02}}{2\Gamma({3-d\02})\0d}\Bigl\{
\phi_B^2(x)\[\({3/4}\)^{(d-1)\02}f(d)-1\]\nn
&&\qquad\qquad\qquad\quad+\phi_0^2(x)\[ d \({3/4}\)^{(d-1)\02}f(d)-2 \] \Bigr\}
\eea
where
\be
f(d)={}_2F_1\({3-d\02},{1\02};{3\02};-{1\03}\)
\ee
with the special cases
\bea
f(1)&=&\frac{\pi }{2 \sqrt{3}},\qquad
f(2)=\frac{1}{2} \sqrt{3} \ln3,\nn
f(3-2\varepsilon)&=&1+\varepsilon\(
2-\frac{\pi }{\sqrt{3}}-\ln\left(\frac{4}{3}\right)
\).
\eea
These can be used to write out closed-form results for
the local Casimir energy density in 1+1 dimensions (calculated previously
in Ref.~\cite{Goldhaber:2001rp}) and for the 2+1-dimensional
domain wall.
With $\int dx\, \phi_0^2=\int dx\, \phi_B^2=1$ one can readily verify
that the {\em total} energy obtained from $\int dx\, \epsilon_{\rm Cas}^{(T=0)}(x)$
agrees with the result given in Eq.~(19) of Ref.~\cite{Rebhan:2002uk},
which is finite for all $d\le4$.

But for the {\em local} energy density
we have UV divergent contributions for $d=3-2\varepsilon$,  
\be\label{epsCasdiv}
\epsilon_{\rm Cas}^{{\rm div.}}(x)|_{d=3-2\varepsilon}=
{1\06\varepsilon}{m^3\0(4\pi)^2}\[\phi_0^2(x)-\phi_B^2(x)\],
\ee
which turns out to cancel incompletely with the UV divergence
found above in Eq.~(\ref{DepsCasphi1}),
\be\label{epsDCasdiv}
\Delta\epsilon_{\rm Cas}^{{\rm div.}}(x)|_{d=3-2\varepsilon}=
-{1\02\varepsilon}{m^3\0(4\pi)^2}\[\phi_0^2(x)-\phi_B^2(x)\].
\ee

However, it is well known that the local energy density is ambiguous
up to improvement terms to the energy-momentum tensor, and that
an unimproved energy-momentum tensor in general needs composite
operator renormalization \cite{Callan:1970ze,Freedman:1974gs,Freedman:1974ze,Collins:1976vm,Brown:1980pq}. 
This ambiguity signals that the local energy density is not
a well-defined physical observable as such. If one defines the
energy density as the 00-component of the 
gravitational energy-momentum tensor (variation
of the Lagrangian with respect to the metric)
then it will depend on the particular coupling to the gravitational
field, which can also include a term in the Lagrangian
of the form  $-\2\xi\sqrt{-g(x)}R(x)\varphi(x)^2$ 
where $R$ is the Riemann scalar. In a flat spacetime, this
produces an additional contribution to the local energy density given by
\be
\epsilon_\xi(x)=-\xi\6_x^2\varphi^2(x),
\ee
which can contribute already at tree-level for a ``nonminimal'' coupling
$\xi\not=0$.\footnote{The associated integrated total energy vanishes for the
kink, but in solitons involving long-range fields such as magnetic
monopoles improvement terms to the energy-momentum tensor can also
contribute to the total energy, see Ref.~\cite{Rebhan:2005yi}.}
 Even when $\xi=0$ at tree level, as we have
implicitly been assuming so far, quantum contributions will in general
produce such a term. Indeed, the combination $\phi_0^2-\phi_B^2$ appearing
in the divergent contributions Eqs.~(\ref{epsCasdiv}) and (\ref{epsDCasdiv})
is proportional to $\6_x^2\varphi_K^2(x)$,
\be
\phi_0^2-\phi_B^2={6\lambda\0m^3}\6_x^2\varphi_K^2(x)
\ee
and thus can be removed by a counterterm $-\2\delta\xi\sqrt{-g(x)}R(x)\varphi(x)^2$ 
with divergent $\delta\xi$.

Alternatively, having
a nonzero $\xi$ already at tree-level produces further divergences of
the form of (\ref{epsCasdiv}) and (\ref{epsDCasdiv}), namely
\be
\Delta\epsilon_\xi=-\xi\6_x^2\<\eta^2\>(x)=-4\xi \Delta\epsilon_{\rm Cas}.
\ee
This is UV divergent in 3+1 dimensions and also generally IR divergent because
it involves the kink profile. By choosing $\xi=1/2$ we can
in fact make sure that the correspondingly modified local energy density
is independent of $A_d(T)$ and thus always IR finite (but not UV finite
in 3+1 dimensions). 
Alternatively, $\xi=1/6$ leads 
to a UV finite (but generally IR divergent) one-loop local energy density for $d=3$
without infinite renormalization
of $\xi$. 
(In a massless theory, $\xi={d-1\04d}$ preserves conformal invariance
in a gravitational background and thus $\xi=+1/6$ is the value which
corresponds to an improved stress tensor that is finite in 3+1 dimensions.)

\section{Conclusion}

\begin{table}
\caption{\label{tab:table1}Divergences left after standard
renormalization
in the field profile $\phi_1(x)$, the local energy density $\epsilon(x)$,
and total energy (mass or surface tension)
for the various dimensions. ``IR'' denotes IR divergences which
introduce a dependence on system size; ``UV'' denotes divergences
requiring composite operator renormalization through
improvement terms for the energy momentum tensor.}
\begin{ruledtabular}
\begin{tabular}{rlll}
dimension&profile $\phi_1(x)$&energy density $\epsilon(x)$&total energy $M$\\
\hline
1+1 & finite & finite & finite \\
2+1 & IR (all $T$) & IR (all $T$, $\xi\not=\frac12$) & finite \\
3+1 & IR ($T>0$) & IR ($T>0$, $\xi\not=\frac12$), 
UV ($\xi\not=\frac16$) & finite \\
\end{tabular}
\end{ruledtabular}
\end{table}

We have calculated one-loop corrections to the profile
of sine-Gordon and CP$^1$
kinks and $\varphi^4$ domain walls at zero and finite temperature. 
Using dimensional regularization,
we have reproduced results for the one-loop field profile
of 1+1-dimensional (bosonic) kinks of Ref.~\cite{Shifman:1998zy,Goldhaber:2001rp}, and have extended them to include
thermal contributions, and also to cover higher-dimensional $\varphi^4$
kink domain walls. We have shown that a renormalization condition
which defines thermal screening masses self-consistently at negative
wave vector squared simplifies the result and
removes certain artefacts in the resulting one-loop corrected kink profile.

However, in the case of domain walls, we have encountered
divergences in local quantities that are not taken care of by standard
renormalization of the parameters in the Lagrangian
as summarized in Table~\ref{tab:table1}.
On the one hand, we have found
infrared singularities
caused by the massless modes which are the higher-dimensional analogs of
the translational zero mode of the 1+1-dimensional kink. 
In the 2+1-dimensional
domain wall, these infrared singularities lead to a logarithmic
sensitivity of the correction to the classical kink profile
on the system size. This corresponds to the phenomenon
of interfacial roughening in statistical physics, where the
same model has been studied in a 3-dimensional
Euclidean setting \cite{Kopf:2008hr}. In 3+1 dimensions,
the one-loop kink profile at zero temperature
turns out to be infrared finite in dimensional
regularization, whereas thermal contributions lead to
logarithmic infrared singularities, thus exhibiting the
phenomenon of interface roughening in the context of
a relativistic field theory at finite temperature.

In the case of the $\varphi^4$ kink and the corresponding
domain walls, we have also
calculated the local energy profile and discussed its ambiguities.
Depending on the underlying energy-momentum tensor, in particular
the parameter $\xi$ in a possible $-\2\xi\sqrt{-g}R\varphi^2$ term
in the Lagrangian, we generally found both infrared and ultraviolet
divergences. In 2+1 and 3+1 dimensions, 
the infrared divergences are the same as those found
in the field profile, although for the choice $\xi=\2$ they
can be eliminated from the energy profile.
Ultraviolet divergences arise in 3+1 dimensions
whenever $\xi\not=\frac16$, corresponding
to an unimproved energy-momentum tensor, and these divergences can then be
cancelled by improvement terms.
However, both infrared and ultraviolet divergences drop out in the
integrated energy density.
Considering the manifestly finite thermal corrections to
the energy density we have found that 
the contributions from the massless modes are crucial
in ensuring that the energy density vanishes in the limit $m\to0$, i.e.\
when the domain wall has melted in a second-order phase transition.
In the case of the field profile, excluding
these massless modes (as done e.g.\ in Ref.\ \cite{Graham:2001kz})
would have resulted in non-renormalizable ultraviolet-divergences
in place of the (physical) infrared divergences.

\acknowledgments

We thank Gernot M\"unster and Jan Smit for valuable discussions and
acknowledge financial support from the Austrian Science Foundation FWF, project no.\ P19958.


\end{document}